\documentclass[fleqn,10pt]{wlscirep}
\usepackage{amsmath,amssymb,wasysym}
\usepackage{xcolor}
\newcommand{\red}[1]{\textcolor{black}{#1}}

\usepackage{siunitx}
\usepackage{lineno}
\title{Lagrangian Statistics and Intermittency in Gulf of Mexico}
\author[1]{Liru Lin}
\author[2*]{Wei Zhuang}
\author[2]{Yongxiang Huang}
\affil[1]{South China Sea Institute of Planning and Environmental Research, State Oceanic Administration, Guangzhou, 510300, PR China}
\affil[2]{State Key Laboratory of Marine Environmental Science \& College of Ocean and Earth Sciences,
Xiamen University, Xiamen 361102, PR China}
\affil[*]{wzhuang@xmu.edu.cn}

\begin{abstract}
Due to the nonlinear interaction between different flow patterns, for instance, ocean current, meso-scale eddies, waves, etc, the movement of ocean is extremely complex, where  a multiscale statistics is then relevant.  In this work, a  high time-resolution velocity with a time step 15 minutes  obtained by the Lagrangian drifter deployed in the Gulf of Mexico (GoM) from July 2012 to October 2012  is considered. The measured Lagrangian velocity correlation function shows a strong daily cycle due to the  diurnal tidal cycle.
 The estimated Fourier power spectrum $E(f)$ implies  a dual-power-law behavior which is separated by the daily cycle. The corresponding scaling exponents are close to $-1.75$ and $-2.75$ respectively for the time scale larger (resp. $0.1\le f\le 0.4\,\si{day^{-1}}$) and smaller (resp. $2\le f\le 8\,\si{day^{-1}}$) than 1 day.  A 
Hilbert-based approach is then applied to this data set to identify the possible multifractal  property of the cascade process.  The results show an intermittent dynamics for the time scale larger than 1 day, while a less intermittent dynamics for the time scale smaller than 1 day.   It is speculated that the energy is partially injected via the diurnal tidal movement and then transferred to larger and small scales through a complex cascade process, which needs more studies in the near future. 
\end{abstract}
\begin{document}

\flushbottom
\maketitle
\thispagestyle{empty}

\section*{Introduction}
\begin{figure}[!htb]
\centering
\includegraphics[width=0.65\linewidth,clip]{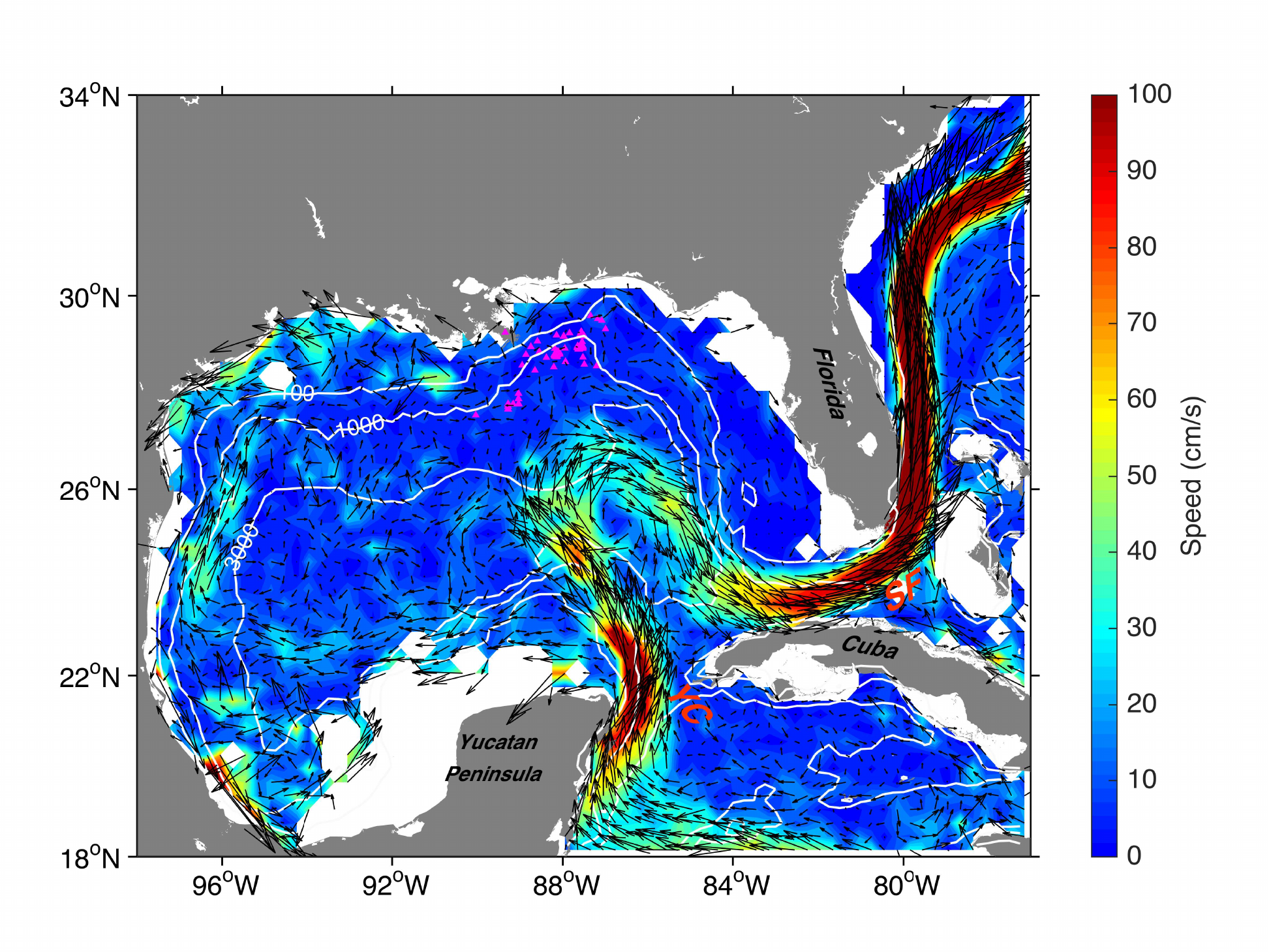}
  \caption{(Color online) \red{Mean sea surface current in the Gulf of Mexico averaged within $0.3\times0.3$ bin from the 2000-2016 Global Drifter Program trajectory dataset. White 
  lines show the $100$, $1000$ and $3000$ m isobaths. Purple triangles indicate the initial 
  positions of the drifters deployed during the GLAD program. YC and FS represent Yucatan 
  Channel and Florida Straits respectively. Figure is plotted using MATLAB R2014b (http://
  www.mathworks.com/) with the M\underline{\hspace{0.2cm}}Map (a mapping package, http://www.eos.ubc.ca/$\sim$rich/map.html).} }\label{fig:Mexico}
\end{figure}

The movement of the ocean is extremely complex due to the nonlinear interaction between different flow patterns, where turbulence may play an important role.\cite{Thorpe2005book}  For instance, the energy could injected to the system via the instability of the ocean current with a length scale hundreds or thousands kilometers and then transferred to the so-called mesoscale eddies through a possible cascade process. A better understanding of this process is crucial for not only the ocean dynamics, but also an ideal testbed with high Reynolds numbers for turbulence theory.\cite{Tennekes1972}  The Gulf of Mexico (GoM) is such a typical region exhibiting a very complex dynamics, such as Loop Current (LC), diurnal tide, mesoscale and sub-mesoscale eddies, etc, \red{see an illustration in 
Fig.\,\ref{fig:Mexico}}. It is a semi-enclosed marginal sea located west of the Atlantic Ocean, connected with the Atlantic Ocean to the east via the Straits of Florida and the Caribbean Sea to the south via the Yucatan Channel. The GoM circulation is characterized by strong current possessing notable variability. The LC is the most energetic components of ocean circulation in the GoM and significantly affect multi-scale processes herein. It originates from the northward-flowing Yucatan Current. After passing through the Yucatan Channel, the LC circulates anticyclonically in the eastern GoM and then exits through the Straits of Florida \cite{Forristall1992JGR,Oey2005}. Within the GoM, the LC displays a wide range of spatiotemporal variability and episodically sheds anticyclonic rings, which are $\sim300\,\si{km}$ in diameter and $\sim 1000\,\si{m }$ in vertical extent. \cite{Forristall1992JGR,Oey2005,Donohue2016DAO,Liu2016JGR,Elliott1982JPO}. The time interval between the ring shedding events varies from a few weeks to $19$ months \cite{Vukovich1995JGR,Sturges2000JPO,Leben2005} with a mean period of about $8$ months.\cite{Dukhovskoy2015DSR,Lugo2016DAO} Besides the large warm-core rings, relatively smaller-scale frontal eddies and filaments are also observed around the edges of LC and its rings by in-situ and remote sensing data, indicating the active mesoscale and sub-mesoscale variability in the GoM. \cite{Vukovich1985JPO,Zavala2003JPO,Henaff2014PiO} Meanwhile, the LC's impact could also extends to the deep ocean, exciting topographic waves and bottom-intensified cyclonic eddies beneath the anticyclonic rings. \cite{Hofmann1986JGR,Hamilton2009PiO,Cherubin2006JPO} Numerical studies indicates that the LC-topography interactions and ring shedding are both in favor of the formation and development of cyclonic eddies, during which cyclones primarily gain energy from LC as a consequence of mean-to-eddy energy conversion.\cite{Donohue2016DAOB,Cherubin2006JPO,Oey2008JPO}  The northeastern GoM is characterized by complex bathymetry, with a right-angle submarine valley, named the DeSoto Canyon, between two wide shelves (the West Florida Shelf and the Mississippi-Alabama Shelf). In this region, the local winds, eddy activities, topographic waves and Mississippi River input could jointly influence the interplay between the shelf and deep circulations, thus resulting in notable cross-shelf exchanges. \cite{Wang2003JPO,Ohlmann2005PiO,Weisberg2005,Hamilton2015CSR} The LC rarely extends sufficiently northward to the DeSoto Canyon region. But it exerts indirect impacts on the shelf-slope flows around this region through either its associated eddies or the coastal-trapped waves exited by its impingement on the West Florida Slope.\cite{Hamilton2015CSR,Hallock2009JPO,Nguyen2015GRL}  The multiscale or scaling property of the ocean movement in GoM region is seldom been investigated. For example, the influence of the sub-mesoscale on two-drifter dispersion has been studied, where the Richardson-Obukhov scaling has been reported for the GLAD (Grand LAgrangian Deployment) experiment \cite{Poje2014PNAS}.
A Kolmogorov-like scaling in space for the second-order Eulerian structure-function  is obtained from the same Lagrangian drifter experiment  \cite{Poje2017PoF}.

\section*{Results}

\begin{figure}[!htb]
\centering
\includegraphics[width=0.65\linewidth,clip]{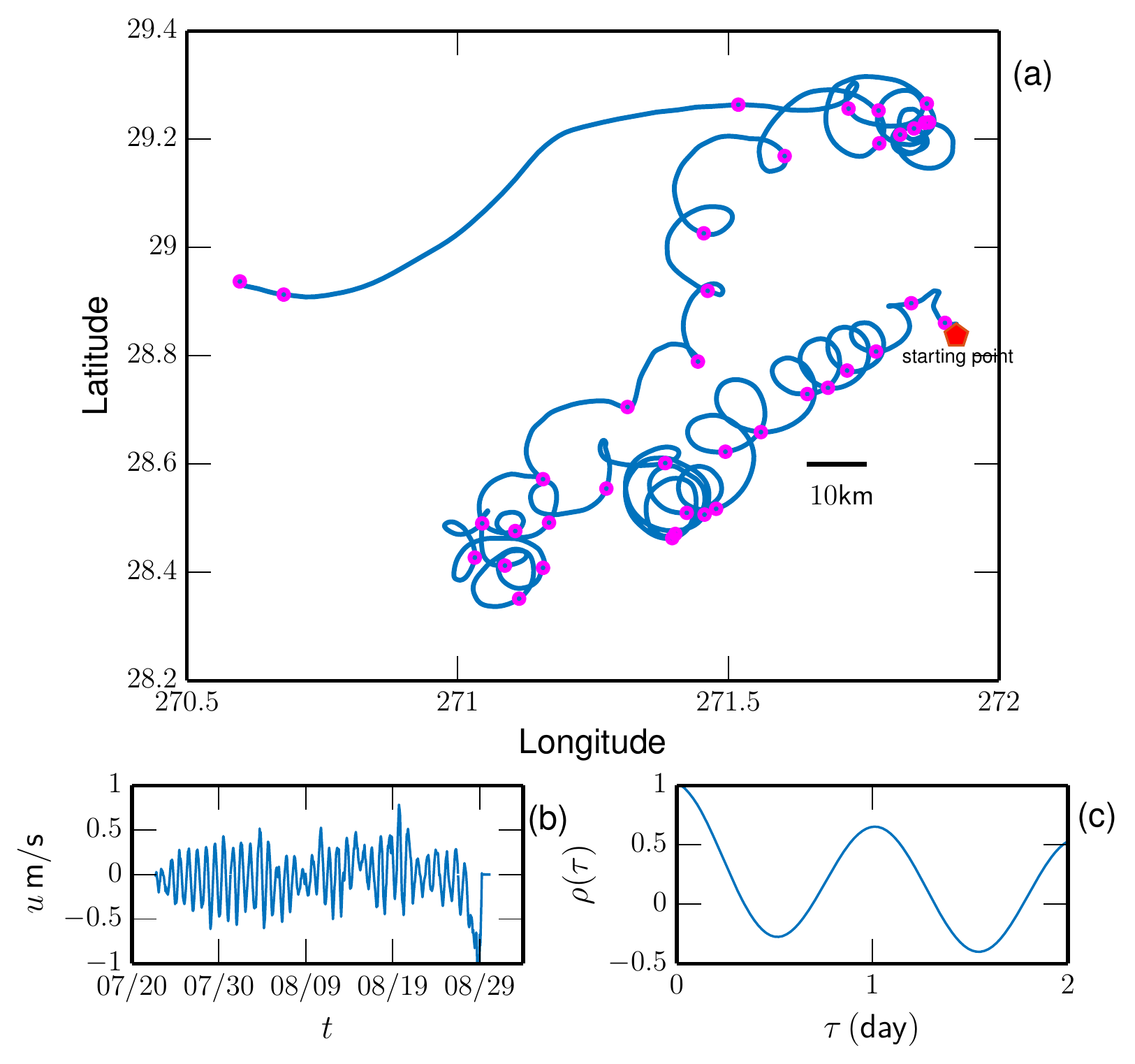}
  \caption{(Color online) (a) Illustration of a 38 day trajectory of Lagrangian drifter on the time period 22th Jul. to 30th Aug. 2012. The 1 day time interval is indicated by symbols. (b) The corresponding Lagrangian velocity. (c) The measured Lagrangian correlation function $\rho(\tau)$, showing a strong daily cycle due to the tide.  This figure is prepared using a Python package, namely Matplotlib. }\label{fig:Trajectory}
\end{figure}

\begin{figure}[!htb]
\centering
\includegraphics[width=0.65\linewidth,clip]{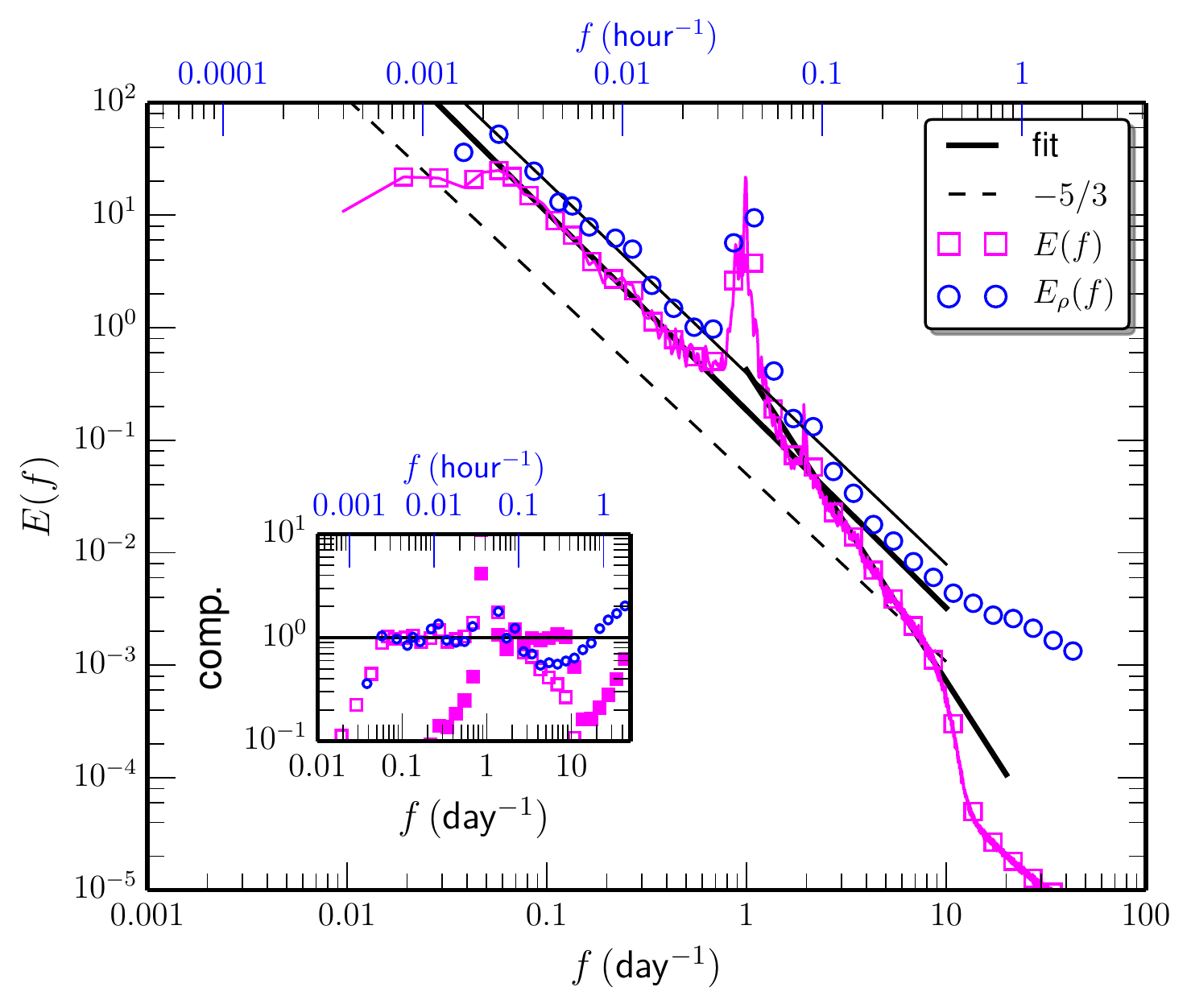}
  \caption{(Color online) Measured Fourier power spectrum $E(f)$ provided by padding zeros and autocorrelation function (denoted as $E_\rho(f)$). Power-law behavior is observed on the range $0.06\le f\le 0.6\,\si{day^{-1}}$, and $2\le f\le 8\,\si{day^{-1}}$ respectively with scaling exponent $\beta_L^F=1.75\pm0.08$ (resp. $\beta_{\rho,L}^F=1.68\pm0.10$) and $\beta_S^F=2.75\pm0.3$.  
For reference, the Kolmogorov $-5/3$ scaling for Eulerian velocity  is shown as dashed line. For display clarity, the spectrum curve has been vertical shifted. 
The inset shows the compensated curve to emphasize the observed power-law behavior.  This figure is prepared using a Python package, namely Matplotlib.}\label{fig:Fourier}
\end{figure}

\subsection*{Autocorrelation function and Fourier power spectrum}
Figure \ref{fig:Trajectory}\,(a) shows a 38 days long  trajectory of Lagrangian drifter, where the symbol indicates the time interval 1 day, and (b) the corresponding zonal velocity $u(t)$.   Graphically, the daily cycle due to the diurnal tide is evidenced.  To show this more clearly, the autocorrelation $\rho(\tau)$ is estimated, see \ref{fig:Trajectory}\,(c), where a strong daily cycle is visible.  
Figure \ref{fig:Fourier} displays  the measured Fourier power spectrum $E(f)$ by padding zeros to have the same 
length for each drifter ($\square$) and by applying the Wiener-Khinchin theorem to measured  autocorrelation function $\rho(\tau)$ (denoted as $E_{\rho}(f)$, $\ocircle$). Power-law behavior is 
observed for $E(f)$, e.g., $E(f)\propto f^{-\beta}$ on the range $0.05\le f\le 0.5\,\si{day^{-1}}$ and $2\le f\le 8\,
\si{day^{-1}}$, corresponding to a time scale range $2\le \tau\le 20\,\si{day}$ and $3\le \tau \le 12\,\si{hour}$. The 
measured scaling exponents are $\beta^F_L=1.75\pm0.08$ and $\beta^F_S=2.75\pm0.30$, \red{where $F$ presents for the Fourier power spectrum, $L$ for the time scale larger than 1 day, and $S$ for the time scale smaller than 1 day.}.  The inset shows the compensated curve to emphasize the observed scaling behavior. 
A statistical test shows that the padding zeros method overestimate the scaling exponent $\beta_L$ (not shown here). 
The second approach detects power-law behavior  for the first scaling  range, i.e., $0.05\le f\le 0.5\,\si{day^{-1}}$ with a scaling exponent $\beta_{L}^{\rho}=1.68\pm0.10$. While the statistical test shows that the second scaling  is biased. 
Both approaches predict a scaling exponent close to the Kolmogorov $-5/3$ for the low frequency part.  Note that the Kolmogorov-Landau theory predicts a power-law behavior $E(f)\sim f^{-2}$ for the three-dimensional homogeneous and isotropic turbulence \cite{Landau1987}.  This type scaling has been reported for zonal and meridional velocity
\cite{Rupolo2007JPO}.
However, due to the existence 
of the strong diurnal tide, the measured  $q$th-order Lagrangian structure-function, e.g., $S_q(\tau)=\langle \vert u(t+
\tau)-u(t)\vert ^q \rangle_t$ fails to detect the corresponding power-law behavior \cite{Poje2017PoF}. Therefore, whether the 
scaling range possess intermittency correction or not can not be distinguished via the conventional approaches, such as structure-function, detrended fluctuation analysis \cite{Schmitt2016Book}.

\begin{figure}[!htb]
\centering
\includegraphics[width=0.65\linewidth,clip]{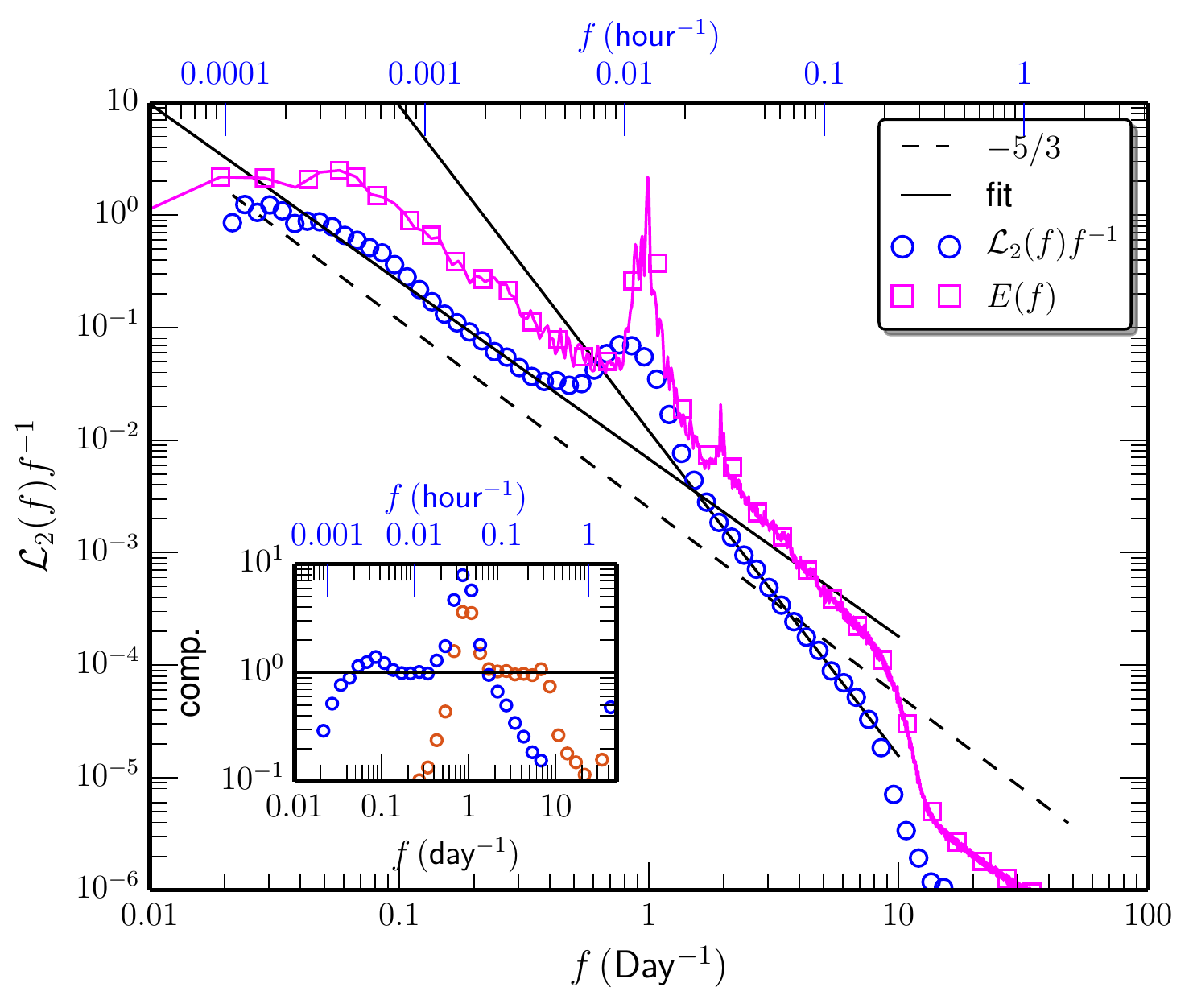}
  \caption{(Color online) Measured Hilbert-based power spectrum $\mathcal{L}_2(f)f^{-1}$ ($\ocircle$). Power-law behavior is observed on the range $0.1\le f\le 0.4\,\si{day^{-1}}$, and $2\le f\le 8\,\si{day^{-1}}$ respectively with scaling exponent $\beta_L^H=1.59\pm0.08$  and $\beta_S^H=2.89\pm0.07$.   For comparison, the Fourier power power spectrum $E(f)$ is also shown as $\square$.
 For display clarity, the spectrum curve has been vertical shifted. 
The inset shows the compensated curve to emphasize the observed power-law behavior.  This figure is prepared using a Python package, namely Matplotlib.}\label{fig:Hilbert}
\end{figure}

\begin{figure}[!htb]
\centering
\includegraphics[width=0.65\linewidth,clip]{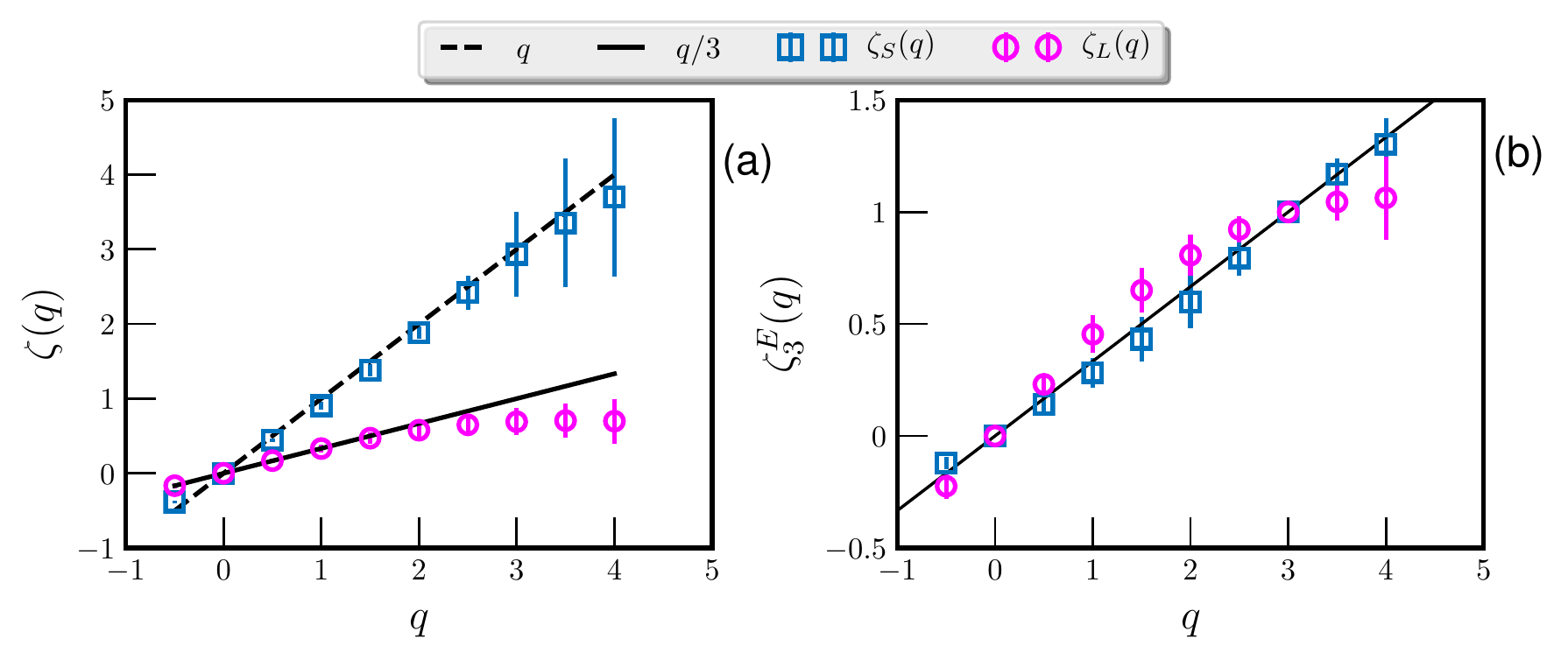}
  \caption{(Color online) (a) Measured Hilbert-based scaling exponent $\zeta_S(q)$ ($\square$) and $\zeta_L(q)$ ($\ocircle$) respectively for   high and low-frequency scaling ranges.  For reference, $\zeta(q)=q/3$ and $\zeta(q)=q$ are shown as solid and dashed lines.  (b) The relative scaling exponent $\zeta^E_3(q)$ by fitting the measured $\mathcal{L}_q(f)$ versus $\mathcal{L}_3(f)$.  This figure is prepared using a Python package, namely Matplotlib.}\label{fig:Scaling}
\end{figure}
\subsection*{Hilbert Statistics and Intermittency Corrections}

Figure \ref{fig:Hilbert}  shows the measured Hilbert-based energy spectrum $\mathcal{L}_2(f)f^{-1}$ ($\ocircle$), where the Fourier power spectrum $E(f)$ ($\square$) is also shown for comparison.  Power-law behavior is observed on the range $0.1\le f\le0.4\,\si{day^{-1}}$ and $2\le f\le 8\,\si{day^{-1}}$ with scaling exponent $\beta_L^H=1.59\pm0.08$ and $\beta_S^H=2.89\pm0.07$,\red{where $H$ presents for the Hilbert-based approach}. Note that there is no half-day harmonic in the Hilbert curve since the Hilbert-based does not require  harmonic to mimic the nonlinear process \cite{Huang1998EMD}. Meanwhile, the scaling exponent $\beta_L^H$ is smaller than the one predicted by the Fourier analysis, which is an effect of  the finite size sample. The measured $\beta_S^H$ is on the same level as $\beta_S^F$ provided by Fourier analysis.  

The high-order scaling exponents $\zeta(q)$ are then calculated on the same scaling range for the $q$th-order Hilbert-based moments $\mathcal{L}_q(f)$ with $q$ on the range $-0.5\le q\le 4$. Figure \ref{fig:Scaling}\,(a) shows the measured $\zeta(q)$, where the value $\zeta(q)=q/3$ (solid line) for the Kolmogorov's 1941 scaling and $\zeta(q)=q$ (dashed line) are illustrated for comparison. Visually, the measured $\zeta_L(q)$ is convex and deviates from $q/3$ when $q\ge 2$, indicating an energy-like scaling with an intermittency correction. Moreover, $\zeta_S(q)$ agrees well with $q$, implying an enstrophy-like scaling with a less intermittency correction.  To compare the potential intermittency with the same reference line, the relative scaling exponent $\zeta^E_3(q)$  by fitting the measured $\mathcal{L}_q(f)$ versus $\mathcal{L}_3(f)$ on the scaling range. The measured $\zeta_3^E(q)$ is displayed in Fig.\ref{fig:Scaling}\,(b).  It confirms that the scaling behavior in the low frequency part is intermittent, while the high frequency one is less intermittent.

\section*{Possible Cascade Dynamics}
A weak stratification  with depth of $10\sim15\,\si{m}$ is reported\cite{Poje2017PoF}.  The flow topography is thus quasi-2D. The Kraichnan's 2D turbulence picture \cite{Kraichnan1967PoF} could be applied here with more complex conditions: the energy is partially injected into the system via  the strong diurnal tide with a typical time scale $1\,\si{day}$.  It is then transferred to high frequency part via a forward cascade, and transferred to low frequency part via an inverse cascade in the Lagrangian point of view. However, due to the complexity of the problem, this simple turbulence theory cannot be applied here directly to predict the scaling exponent.
\red{Another possible  interpretation  could be based on the so-called geostrophic turbulence\cite{Charney1971JAS}, where the energy is injected into the system mainly via an instability of the large-scale circulation and then transferred to small scales \cite{Vallgren2011PRL}. In this situation, the scaling exponent for the large- and small-scale parts are respectively $-3$ and $-5/3$, which is on the opposite of our observation. However, to exclude any theory, a more detail examination of the scale-to-scale  energy flux\cite{Zhou2015JFM} is required to determine the direction of the cascade.}
 It opens a new challenge to theoreticians to propose new turbulence theory in the Lagrangian frame by taking into account more facts, such as diurnal tide, stratification, earth rotation, ocean current, etc. 

\section*{Methods}
\subsection*{Lagrangian Drifter  Data in Gulf of Mexico}
The Lagrangian drifter data is collected during the GLAD (Grand LAgrangian Deployment) observational program in the DeSoto Canyon region of the northern Gulf of Mexico. The experiment was conducted  from July to October 2012 with approximately 300 standard CODE surface drifters were released over a two week period. The CODE GPS-tracked drifter is designed to follow currents in the upper 1 \si{m} with $1-3\,\si{cm/s}$ velocity errors for wind speeds up to $10\,\si{m/s}$. \cite{Poje2017PoF} The mean  life-time of the drifter is around $\sim 56\,\si{days}$ with a standard deviation around $\sim27\,\si{days}$. According to the collected data, a stratification with $10\sim15\,\si{m}$ is observed. The same data set has been analyzed for the sub-mesoscale motion in GoM  \cite{Poje2017PoF,Poje2014PNAS}. The GLAD drifter data used in this work is publicly available \footnote{\"Ozg\"okmen T (2012) CARTHE: GLAD experiment CODE-style drifter trajectories 
(low-pass filtered, 15 minute interval records), northern Gulf of Mexico near DeSoto Canyon, July-October 2012. Gulf of Mexico Research Initiative, 10.7266/N7VD6WC8. Available at https://data.gulfresearchinitiative.org/data/R1.x134.073:0004.}
\red{In this study, all these 300 drifters are considered.}

\subsection*{Autocorrelation Function and Fourier Power Spectrum}
It is often that the collected data is with different sample size for different realizations.  It leads difficulty in calculating some statistical quantities, for instance, Fourier power spectrum. In this work, two different approaches are considered. 
The first one is to extend the data set to have the same sample size by padding zeros to the end of the collected data. 
A numerical experiment shows that  low frequency part will be slightly biased this approach, for example the scaling exponent $\beta_L^F$ is slightly overestimated. 

The second approach is based on the Wiener-Khinchin theorem. 
The autocorrelation function is firstly estimated as,
\begin{equation}
\rho(\tau)=\frac{1}{M(\tau)}\sum_{i=1}^N \tilde{u}_i(t+\tau)\tilde{u}_i(t)  %\frac{\langle \tilde{u}(t+\tau)\tilde{u}(t)  \rangle_t}{\sigma^2}
\end{equation}
where  $\tilde{u}(t)=u(t)-\langle u(t) \rangle_t $ is the fluctuation velocity;   $\tau$ is the time lag; $N$ is the number of drifters and $M(\tau)$ is the sample size for time lag $\tau$. According to the  the Wiener-Khinchin theorem, the corresponding Fourier power spectrum can be estimated via the Fourier transform,
\begin{equation}
E(f)=\int_{-\infty}^{+\infty}\rho(\tau)\cos(2\pi f\tau) d \tau
\end{equation}
where $f$ is frequency.  A numerical experiment shows that the high-frequency part will be biased due to the different length of trajectories, which could be suppressed via a systematically way (not employed here).

\subsection*{Hilbert-Huang Transform}
To constrain the influence of the daily cycle, we employ here  the so-called Hilbert-Huang transform (HHT), which is introduced by N.E Huang \cite{Huang1998EMD}.  The first step of this methodology is to decompose a given velocity $u(t)$ data into a sum of intrinsic mode functions (IMFs) $C_i(t)$ via the so-called empirical mode decomposition (EMD) algorithm without  \textit{a priori} basis functions \cite{Huang1998EMD}. An IMF has to satisfy the following conditions: (i) in the whole data set, the number of extrema and the number of zero-crossings must either or differ at most by one and (ii) the mean value of the envelope defined by the local maxima and the envelope defined by the local minima is zero.  
The IMF is thus a pure oscillatory mode bearing amplitude and frequency modulations that can be extracted by the Hilbert spectral analysis \cite{Huang1998EMD,Schmitt2016Book} as following,
 \begin{equation}
 \tilde{C}^A(t)= C_i(t)+j\frac{1}{\pi} P\int\frac{C_i(t')}{t-t'} dt'=\mathcal{A}_i(t)\exp(j\phi_i(t))\label{eq:Hilbert}
 \end{equation}
 where $C_i(t)$ is the extracted IMF; $j=\sqrt{-1}$;  $P$ means Cauchy principle value; $\mathcal{A}_i(t)$ and $\phi_i(t)$ are amplitude function and phase function, respectively. The corresponding instantaneous frequency is then defined as, 
 \begin{equation}
 f_i(t)=\frac{1}{2\pi} \frac{d \phi_i(t)}{d t}\label{eq:frequency}
 \end{equation}
 With extracted instantaneous frequency, one can design a $f$-conditioned statistics, 
 \begin{equation}
 \mathcal{L}_q(f)=\left\langle \sum_{i} \vert C_i(t)\vert^q\vert_{f_i(t)=f} \right\rangle_t
 \end{equation}
 For a scaling process, one has a power-law behavior of $\mathcal{L}_q(f)$ 
 \begin{equation}
 \mathcal{L}_q(f)\propto f^{-\zeta(q)}
\end{equation}
where $\zeta(q)$ is the scaling exponent \cite{Huang2008EPL}. 
\red{Note that in this approach, the singularity transform is applied to define the analytical signal (resp. Eq.\,\ref{eq:Hilbert}). Moreover, the first-order derivation of the phase function (resp. Eq.\,\ref{eq:frequency}) is used to define the instantaneous frequency. These two steps have very local ability\cite{Huang1998EMD,Schmitt2016Book}.
This Hilbert-based approach thus can isolate  the influence of  energetic structures \cite{Huang2010PRE,Schmitt2016Book}, such as daily cycle shown here. }

\subsection*{Extended-Self-Similarity}
In ESS, the high-order moments are represented as a function of $p$th-order one in the power-law range to measure the scaling exponent more accurate\cite{Benzi1993PRE}, which is written as
\begin{equation}
\mathcal{L}_q(f)\sim \mathcal{L}_p(f)^{\zeta^\mathrm{E}_p(q)}
\end{equation}
where $\zeta^E_p(q)$ is a relative scaling exponent. For comparison conveniency, we consider here the third-order relative scaling exponent $\zeta^\mathrm{E}_3(q)$ for both large and small scale parts.

%\clearpage
%\newpage
%\bibliography{all}

\section*{Acknowledgements}

 This work is sponsored by the National Natural Science Foundation of China (under Grant Nos. 11332006 and  and 11732010), and  partially by the Sino-French (NSFC-CNRS) joint research project (No. 11611130099, NSFC China, and PRC 2016-2018 LATUMAR ``Turbulence lagrangienne: \'etudes num\'eriques et applications environnementales marines",  CNRS, France).  Y.H. is also supported by the Fundamental Research Funds for the Central Universities (Grant No. 20720150075).

\section*{Author contributions statement}

W.Z. and Y.H. designed this study and analyzed the data.  All authors discussed the physics and contributed to the writing of the manuscript.

\section*{Additional information}
Competing financial interests: The authors declare no competing financial interests.

\end{document}